\documentclass[prd,amsfonts,amsmath,twocolumn,superscriptaddress,nofootinbib]{revtex4}

\usepackage{color}
\usepackage{graphicx} 
\usepackage{epstopdf}
\usepackage{afterpage}

\usepackage{mathpazo}   

\newcommand {\be} {\begin{equation}} 
\newcommand {\ba}{\begin{eqnarray}} 
\newcommand {\ee} {\end{equation}} 
\newcommand{\ea} {\end{eqnarray}}

\begin{document}

\title{Positron production scenarios and the angular profile of the galactic center 511-keV line}

\author{Zainul Abidin}
\affiliation{Department of Physics, College of William and Mary, Williamsburg, VA 23187, USA}

\author{Andrei Afanasev}
\affiliation{Department of Physics, Hampton University, Hampton, VA 23668, USA}
\affiliation{Theory Center, Thomas Jefferson National Accelerator Facility, 
Newport News, VA 23606, USA}

\author{Carl E.\ Carlson}
\affiliation{Department of Physics, College of William and Mary, Williamsburg, VA 23187, USA}

\date{\today}

\begin{abstract}

The observed angular profile of the 511-keV photon excess from the Milky Way galactic center can allow us to select among combinations of various dark matter and other positron production mechanisms with various models for the dark matter distribution.  We find that a relic decay scenario gives too flat an angular distribution for any dark matter distribution in our survey, but that a dark matter-dark matter collisional scenario, or a scenario that involves particles emitted from a localized central source producing positrons some distance out,  can match the observed galactic center angular profile if the dark matter distribution is neither too flat nor too cuspy.  Additionally, positron migration or diffusion before annihilation broadens the angular profile to an extent that an average migration of more than half a kiloparsec is not viable with most dark matter distributions.   The observed angular profile is also consistent with the occurrence of transient events in the past, followed by isotropic positron diffusion.

\end{abstract}

\maketitle


\section{Introduction and Preliminary Results}			\label{sec:one}


The center of our galaxy is observed to be a concentrated source of 511-keV gamma-rays.  The source region appears coincident with the galactic bulge, with a near circular symmetry with about a $6^\circ$ FWHM or of scale about 1 kpc in FWHM diameter and with a total flux from the bulge of about $0.89\times 10^{-3}$ photons cm$^{-2}$ s$^{-1}$~\cite{Knodlseder:2005yq,Weidenspointner:2007rs}.  The identity of the positron source responsible for these gamma-rays remains undetermined.  There have been a number of astrophysical suggestions; some of these suggestions are included in~\cite{Parizot:2004ph,Bertone:2004ek,Guessoum:2006fs,Bandyopadhyay:2008ts,Lingenfelter:2009kx}.  In addition, the possibility that dark matter may create these positrons has been widely discussed~\cite{Pospelov:2007xh,Finkbeiner:2007kk,ArkaniHamed:2008qn,Chang:2008gd,Khlopov:2008ty}. 

In this paper, we focus on the dark matter possibilities and consider what the observed angular profile of the 511-keV gamma-rays tell us about possible dark matter production mechanisms, and further consider whether the dark matter production scenario can inform us regarding the dark matter distribution near the center of the galaxy.  

For purposes of this paper we think of the surviving dark matter particles as WIMPs and will categorize dark matter production as being either from decay of relic long lived excited WIMPs~\cite{Bertone:2007aw,Pospelov:2007xh}, from collisions of WIMPs where the cross section time velocity has no velocity dependence, or from collisions of WIMPs where the cross section has noticeable velocity dependence.  The relic decay idea is that there are excited WIMPs with a small mass gap from the ground state WIMP and which produce positrons when they decay to the lower state~\cite{Finkbeiner:2007kk}.  The lifetime of the excited WIMP needs to be long enough to allow many of them to survive since freeze-out, yet short enough there would be enough decays at present to account for the observed rate, even after considering that only a fraction of the surviving WIMPs would be the excited ones.

The collisional mechanism depends on having final states that lead directly or by some chain to positrons.  One example is WIMP pairs annihilating to lighter dark matter gauge boson pairs, where the dark bosons have a weak coupling to the visible sector so that they soon enough decay into $e^+ e^-$ pairs, 
$\chi \chi \to \phi\phi \to e^+ e^- e^+ e^-$.  This in only an example.  For velocity independent collisional case the spatial distribution of positron production does not depend on the reaction specifics. We only need that the cross section and dark matter density suffice for obtaining the overall rate.

We also consider the possibility that there is a localized source of initiating particles that spread out and at some distance interact in reactions that produce positrons.  Photons streaming outward clearly bremsstrahlung positron-electron pairs, but one can imagine dark matter alternatives.  As a possibility, if WIMPs connect to the luminous world via gauge particles (``dark photons''), one could produce $e^+ e^-$ pairs via a Bethe-Heitler-like process where the WIMP plays the role of the nucleon and the dark gauge boson plays the role of the virtual photon that couples to the nucleon~\cite{Abidin2010b}.  The example is just an example; the angular profile generated by any mechanism where a central source produces particles that then produce positrons some distance out will be the same.  Since the visible galactic bulge is somewhat oblate, seeing a circularly symmetric positron hot spot in the center of the galaxy can be a clue that the dark matter halo is taking a strong role in the production mechanism.  

That the angular (or spatial) profile of the 511-keV gamma-rays can discriminate among dark matter mechanisms is known~\cite{Bertone:2007aw,PalomaresRuiz:2010pn,Chen:2009av,Cline:2010fq}.  The simple reason is that the total number of decays per unit volume in the decay mechanism depends on the dark matter density linearly, while the number of collisions per unit volume depends on the same density quadratically, so that as the observational line of sight gets closer to the center of a dark matter concentration, the collisional mechanism will peak more sharply that the decay mechanism.

In formulas, for the relic decay scenario the angular flux is given by the line of sight integral
\be
\frac{d\Phi}{d\Omega} = \frac{ f^* \Gamma}{2\pi } \int dl  \   n(r)	\,,
\ee
where flux $\Phi$ is the number of 511-keV photons per cm$^2$ per sec, 
$d\Phi/d\Omega$ is per unit solid angle of source, $n(r)$ is the number density of WIMPs, $\Gamma$ is the decay rate of the excited WIMPs, and $f^*$ is the fraction of WIMPs in the excited state.  (The formula is for the case of each decay yielding one positron.)  Similarly for collisions,
\be
\frac{d\Phi}{d\Omega} = \frac{ n_{e^+}  \sigma v }{2\pi } 
	\int dl  \   n(r)	\,,
\ee
for the case that $\sigma v$ is constant and $n_{e^+}$ is the number of $e^+$ produced per collision.  

In the scenario with a central source, the isotropic flux of initiating particles falls like the inverse distance-squared from the source. We will at least \textit{pro tem} place the source at the galactic center, so the line-of-sight integral is like the relic decay case with $n(r)$ replaced by $n(r)/r^2$.  This too will give and angular profile steeper than the relic decay scenario.

Additionally, before plotting any curves the angular flux results are smeared with a $3^\circ$ FWHM gaussian since the data we compare to comes from a telescope with a $3^\circ$ spatial resolution~\cite{Knodlseder:2005yq,Weidenspointner:2007rs}.

\begin{figure}[b]
\begin{center}
\includegraphics[width = 3.0 in]{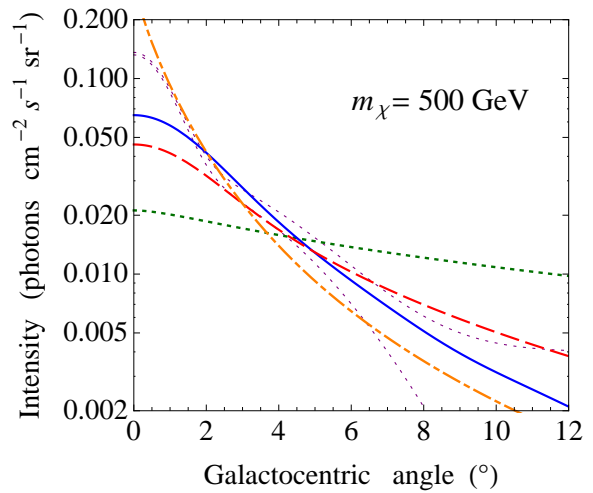}
\caption{Comparing positron production mechanisms and comparing to data. The green dotted line is for the relic decay scenario, the red dashed line is for the velocity independent collisional scenario, the blue solid line is for the collisional scenario with a velocity threshold (the WIMP mass $m_\chi=500$ GeV affects the profile for only this case), and the orange dot-dash curve is for a central source emitting particles that in turn produce positrons at radius $r$.  The calculated curves in this plot use the NFW~\cite{Navarro:1996gj} dark matter distribution. The light dotted curves indicate INTEGRAL data, as explained in the text.}
\label{fig:one}
\end{center}
\end{figure}

The difference between the angular profiles from positrons produced by relic decays and by velocity independent collisions is seen in the heavy dashed (collision) and dotted (decay) curves in Fig.~\ref{fig:one}, using the NFW dark matter distribution model~\cite{Navarro:1996gj}.  We will consider in the next section a selection of other dark matter distributions  and see how the results differ among them.

Also on Fig.~\ref{fig:one} is a heavy solid line for a velocity dependent collision possibility which we shall return to shortly, and two light dotted lines representing the data from the spectrometer SPI on the INTEGRAL (International Gamma Ray Astrophysics Laboratory) satellite.  The central bulge photons have a approximate circular symmetry and there is also a broad 511-keV photon signal from the galactic disk.  The photon distribution from the bulge can be represented by a single gaussian of FWHM $6^\circ$, or somewhat better, by two concentric gaussians of FWHM $2.1^\circ$ and $8.0^\circ$, as~\cite{Weidenspointner:2007rs}
\begin{align}
\frac{d\Phi}{d\Omega} &= \left( 
	0.15 \  \frac{e^{-\theta^2/a_1^2}}{2\pi a_1^2}
	+ 0.74 \  \frac{e^{-\theta^2/a_2^2}}{2\pi a_2^2}
\right)
	\times10^{-3} \, \frac{\rm photons}{\rm cm\ s\ sr}		\,;
\end{align}
for $\theta$ in radians, $a_1 = (2.1\pi/180)/2.355$ and $a_1 = (8.0\pi/180)/2.355$.  (The supplement to~\cite{Weidenspointner:2008zz} suggests somewhat wider gaussians, but~\cite{Weidenspointner:2007rs} states the flux in each gaussian.)  This is shown as the lower light dotted line in Fig.~\ref{fig:one}.  The upper light dotted line is the same but with a constant 0.004 photons cm$^{-2}$ s$^{-1}$
sr$^{-1}$ added to give a idea of the level of the (albeit not circularly symmetric) disk contribution.  The normalization of the data curves is absolute while the normalization of the calculated curves is arbitrary;  we are interested here in the angular shape.  

One sees already that the dark relic positron production curve is too flat to match observation, and this persists even when we consider other models for the dark matter distribution.  Collisional positron production gives an angular profile that falls more steeply and looks closer to the data.

The orange dot-dash curve on Fig.~\ref{fig:one} is for a central source emitting particles isotropically that in turn produce positrons at radius $r$.  One has
\begin{align}
\frac{d\Phi}{d\Omega} = \frac{\sigma_0}{8\pi^2}  \frac{dN}{dt} 
	\int dl \ \frac{n(r)}{r^2} 	\,,
\end{align}
where $dN/dt$ is the emission rate of the initiating particles and $\sigma_0$ is the cross section of the reaction that produces the positrons. 

Finally, the heavy solid curve on Fig.~\ref{fig:one} is from a collisional process with a significant velocity dependence in the cross section.  Typically one models the velocity distribution of dark matter (or other) particles as a Maxwellian with an upper cutoff at the escape velocity~\cite{King:1966fn,Finkbeiner:2007kk}.  As the escape velocity gets higher closer to the center of the galaxy, there are more high velocity particles found there and the rate of collisions near the center is enhanced for this reason as well as for reasons of overall density.  Hence the velocity dependent collisional process rises more steeply toward the center of the galaxy that the velocity independent one, and the match to the data is better.

The specific velocity dependent model we use follows~\cite{Finkbeiner:2007kk} where the velocity dependence occurs because of a threshold.  Velocity dependence can also occur because higher angular momenta are involved, and would lead to qualitatively similar results.  The threshold occurs because, in the model, positron production occurs when an excited WIMP is produced by a WIMP-WIMP collision and the excited WIMP emits  an electron-positron pair as it decays back to the ground state.  If the mass of the WIMP is $m_\chi$ and the mass gap is $\delta$, then the threshold velocity is
\be
v_{thresh}=\sqrt{4\delta/m_\chi}		\,.
\ee

The collisional cross section is modeled as 
\be
\sigma v_{rel}=
\sigma_{0}\sqrt{v_{rel}^2-v_{thresh}^2} \	\ \theta(v_{rel}-v_{thresh})  \,,
\ee
where for WIMP velocities $\vec v$ and $\vec v'$, 
$v_{rel}=|\vec v -\vec v'|$.  
The number of scatterings per time per density is given by
\be
\langle \sigma v \rangle(\vec r)=\int d^3 v \, d^3 v' \, f(\vec v, \vec r) f(\vec v', \vec r) \sigma(v_{rel}) v_{rel}\,,\label{sigmav}
\ee
where $f(\vec v,\vec r)$ is the WIMP velocity distribution at radius $r$.  The calculated flux is modified from the previous result to
\be
\frac{d\Phi}{d\Omega} =  \frac{1}{2\pi} \int dl \ n^2(r)\ \langle\sigma v \rangle (r)\,.
\ee

The velocity distribution is taken to be
\be
f(v,r)=
N(r)  \exp(-v^2/2v^2_{rms}) \ \theta(v_{esc}-v)
\ee
with $v_{rms} = 200$ km/s and normalization $\int d^3v \, f(v,r) = 1$.
The normalization parameter $N(r)$ depends on the distance $r$ from the center of the galaxy because the escape velocity does.

The escape velocity follows from noting the circular rotation velocity $v_c$ of the galaxy is approximately constant out to some radius $r_{max}$.  For the present paper we ignore the variation in the rotation curve near the center of the galaxy. Assuming there is no mass beyond $r_{max}$, one obtains the escape velocity
\be
v_{esc} = 
v_c\sqrt{2[1-\ln(r/r_{max})]}  \,.
\ee
for $r < r_{max}$.  We use $v_c = 220$ km/sec.  The curve in Fig.~\ref{fig:one} is drawn for $r_{max}=25$ kpc. Recent analysis of stellar velocity data suggests that the constant circular speed holds out to 80 kpc~\cite{Gnedin:2010fv}.  Using $r_{max} = 80$ kpc broadens the angular profile somewhat, but not dramatically.

We have thus far only used one dark matter distribution model, and have tacitly (until now) assumed that the positrons annihilate close to where they are produced.  We discuss other dark matter distributions in the next section, and the effects of positron diffusion or migration are discussed in Sec.~\ref{sec:migration}, with just the comment here that positron migration can broaden the angular distributions but not narrow them.

Further, in Sec.~\ref{sec:sporadic}, we discuss possibilities that follow upon the galactic center being more active in the past than it is today.


\section{Dark Matter Distribution Models}


The results in shown in Sec.~\ref{sec:one} used only one model for the dark matter density in our galaxy.  Other dark matter distributions have been suggested, and it is natural to enquire how using these would change the 511-keV angular profile already presented.  

We consider several mass density profiles, in particular, the Merritt \textit{et al.} profile~\cite{Merritt:2005xc}, the Navarro-Frenk-White (NFW) profile~\cite{Navarro:1996gj}, the isothermal halo dark matter profile (ihdp)~\cite{Binney:1995,Ishiwata:2009xg}, a somewhat more singular profile used by Cembranos and Strigari~\cite{Cembranos:2008bw}, here scaled to give the local dark matter density, and a truncated flat (TF) distribution from~\cite{Wilkinson:1999hf} with parameters obtained in~\cite{Battaglia:2005rj}.  The latter halo distribution gives a flat rotation curve all the way to the inner regions of the galaxy, as is observed for the Milky Way at least at radii larger than a few tenths kiloparsec.

\begin{figure}[tp]
\begin{center}
\includegraphics[width = 3.0 in]{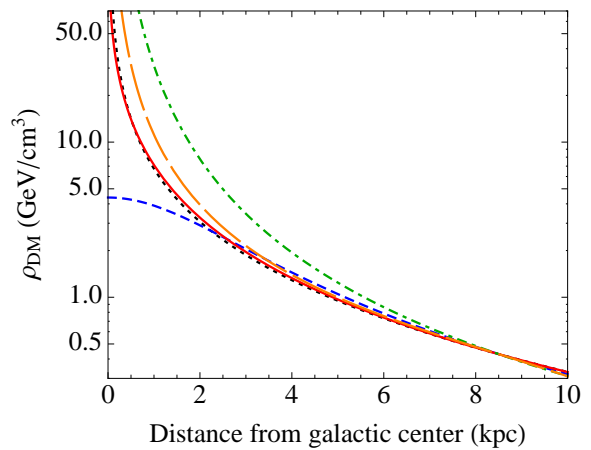}
\caption{Dark matter density models vs.\ distance from the galactic center.  The blue dashed line is the ihdp profile, the solid red line is the Merritt  \textit{et al.} profile, the dotted black line is the NFW profile, the long-dash orange curve is the Cembranos and Strigari profile, and the dot-dash green line is the TF dark matter profile.  The curves are normalized to give the same dark matter density at our sun's location.}
\label{fig:dmprofiles}
\end{center}
\end{figure}

Explicitly, the $\rho_{DM}$ profiles are
\begin{eqnarray}
\rho_{\rm ihdp} &=& \rho_{\rm local} \frac{R_c^2 + R_\odot^2}{R_c^2 + r^2} 
											\nonumber\\
\rho_{\rm NFW} &=& \rho_{\rm local} 
		\frac{R_\odot (1 + R_\odot/r_c)^2}{r (1 + r/r_c)^2}
											\nonumber\\
\rho_{\rm Merritt} &=& \rho_{\rm local} 
		\exp{\left[-\frac{2}{\alpha} 
		\left( \frac{r^\alpha - R_\odot^\alpha}{r_{-2}^\alpha} \right)\right]}
											\nonumber\\
\rho_{\rm Cembranos} &=& \rho_{\rm local}
	\left( \frac{R_\odot}{r}  \right)^{1.5}
	\left( \frac{1 + (R_\odot/r_0)^8}{1 + (r/r_0)^8} \right)^{3/16}
											\nonumber\\
\rho_{\rm TF} &=& \rho_{\rm local}
	\left( \frac{R_\odot}{r}  \right)^2 
	\left( \frac{r^2 + R_\odot^2}{ r^2 + a^2 } \right)^{3/2}
\end{eqnarray}
where $R_c = 2.8$ kpc,  $r_c = 20$ kpc, $\alpha = 0.2$, $r_{-2} = 25$ kpc, $r_0 = 10$ kpc, and $a = 105$ kpc.  We took $R_\odot = 8.5$ kpc and used $\rho_{\rm local} = (0.43 \pm 0.15)$ GeV/cm$^3$ from~\cite{Salucci:2010qr}, with their uncertainties added in quadrature.  (The TF fit parameters from~\cite{Battaglia:2005rj} translate to $\rho(R_\odot) = 0.53$ GeV/cm$^3$, with error limits compatible with $\rho_{\rm local}$ given here; we are in any case interested in this work in the shape of the distribution.) The dark matter profiles give different amounts of dark matter near the galactic center, as can be seen in Fig.~\ref{fig:dmprofiles}.

\begin{figure}[t]
\begin{center}
\includegraphics[height = 1.72 in]{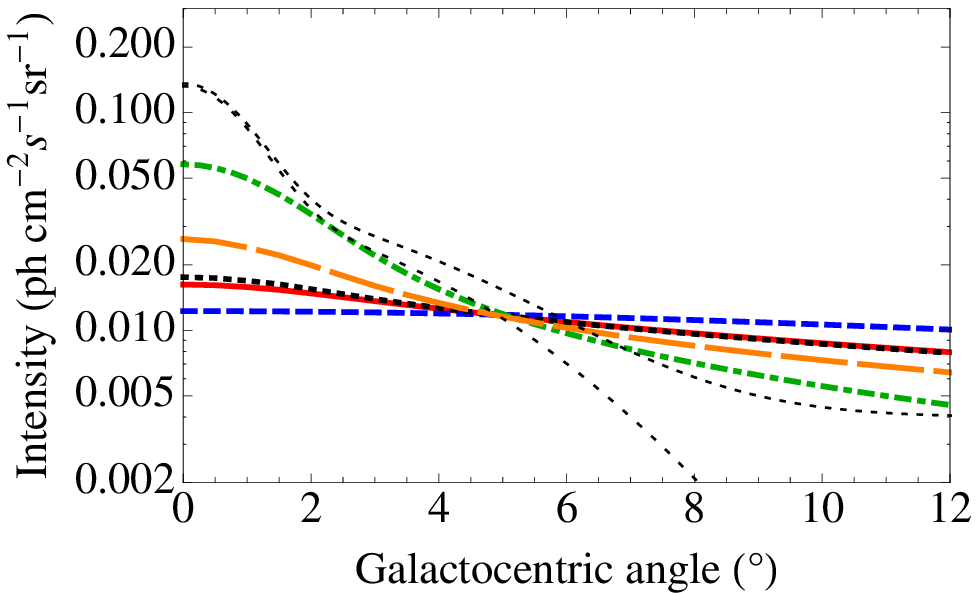}

\includegraphics[height = 1.72 in]{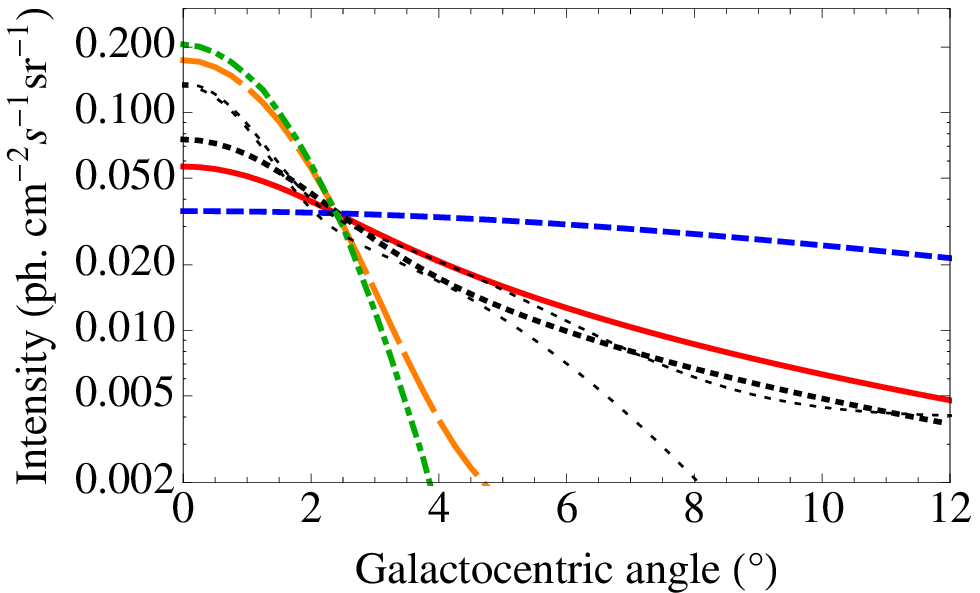}

\includegraphics[height = 1.72 in]{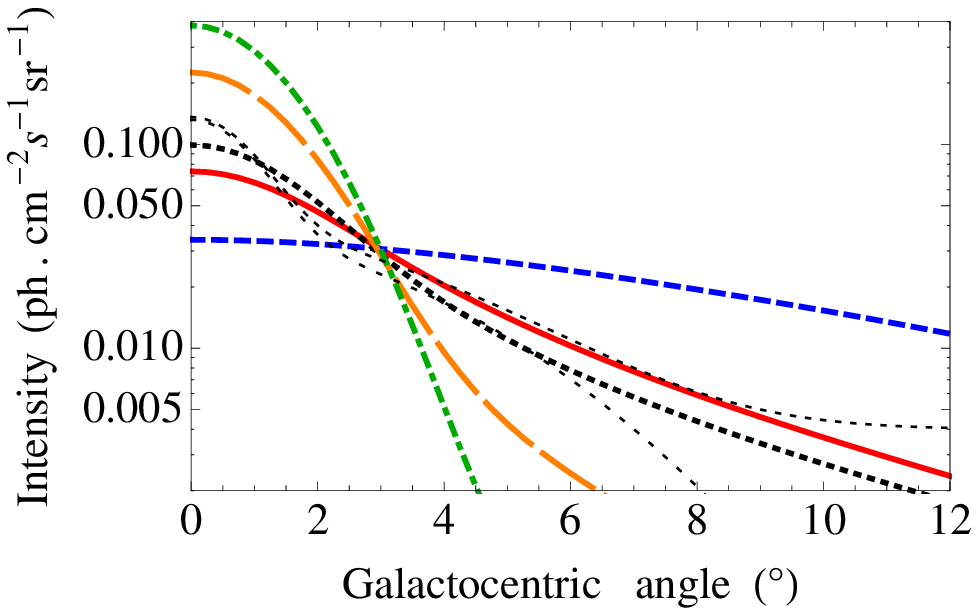}

\includegraphics[height = 1.72 in]{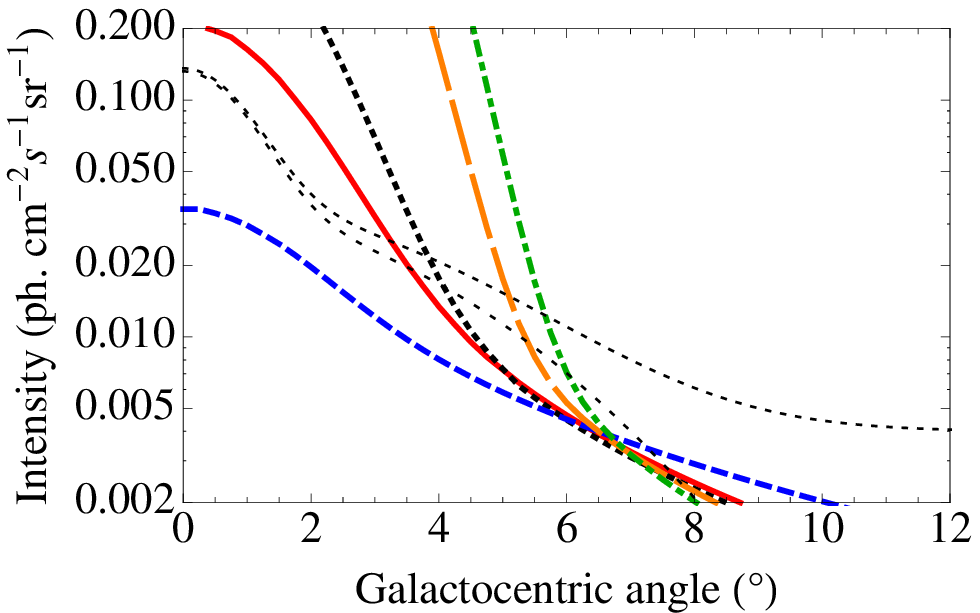}

\caption{Angular distribution of 511-keV photons due to (top) decays of relic excited WIMPs near the galactic center, (second) due to positrons created through WIMP annihilation with $\left<\sigma v\right>$ independent of velocity, (third) due to WIMPs being collisionally excited and subsequently decaying back to the ground state by emitting $e^+ e^-$, and (bottom) due to a point source of initiating particles that spread and produce
positrons when they interact away from the source.  The blue dashed line uses the ihdp profile, the solid red line uses the Merritt \textit{et al.}~profile, the heavy dotted uses the NFW profile, the long-dash orange curve uses the Cembranos and Strigari profile,  and the dot-dash green line uses the truncated flat distribution.  The fainter black dotted lines are a two-Gaussian representation of the data described in the text, one with a $0.004$ photons/(cm$^{2}$sec sr) background included as in~\cite{Finkbeiner:2007kk}.
}
\label{fig:decayprofiles}
\end{center}
\end{figure}

\afterpage{\clearpage}


\subsection{Relic decay positron angular profiles}


The angular distributions of 511-keV photons that follow from the selected dark matter profiles and the relic WIMP decay scenario are shown in Fig.~\ref{fig:decayprofiles}(top).   The normalizations of the curves are controlled by the lifetime of the excited WIMP and by fraction of WIMPs in the excited state.  By choice, our normalizations give results that cross at 5$^\circ$, and assuming each WIMP decay gives one positron, the lifetime/excited WIMP fractions for the displayed results are $\tau/f^* = (4.0,5.1,4.9,6.7,16.1) \times 10^{13}$ years for the ihdp, Merritt, NFW, Cembranos, and TF distributions, respectively.
The light dotted curves in Fig.~\ref{fig:decayprofiles} again represent data from the INTEGRAL satellite~\cite{Knodlseder:2005yq,Weidenspointner:2007rs,Weidenspointner:2008zz}.  


Four of the distributions are too flat to match the data.  The result from the fifth distribution is possible at least in the inner region.  However, this distribution itself ascribes the flatness of the rotation curve to halo matter.  The Milky Way rotation curve is flattish in to even within 1/4 kpc, which is within $1.7^\circ$ from our viewpoint, but usually one thinks that much of the gravitational pull and shape of the rotation curve in the inner region is coming from ordinary matter.  In that case, carrying a $1/r^2$ dark matter profile down to the lowest $r$ overestimates the amount of dark matter there.


\subsection{Velocity independent collision processes}


One can also imagine a scenario of positron production through WIMPs  annihilation. The annihilation occurs without velocity threshold and in the simplest case with $\left<\sigma v\right>$  independent of velocity, hence,  $\left<\sigma v\right>$ can pull out of the line of sight integral. In this case, the angular profile for the 511 keV line photon is flatter than in the excitation case. 

In Fig.~\ref{fig:decayprofiles} (second panel), we show the angular distribution of the 511 keV line for the five different  dark matter profiles. In the plot we use $\langle \sigma v \rangle= 2\pi (1. 7, 0.3, 0.2, 0.01, 0.5\times10^{-6}) \times 10^{-20}$ cm$^3$/s for the ihdp, Merritt, NFW,  Cembranos, and TF distributions, respectively.  The normalizations are chosen so that they cross at $2.5^\circ$.


\subsection{Velocity dependent collision processes}


We calculate the pairs intensity using five different dark matter profiles mentioned above. The WIMP mass is set to $m_\chi=500$~GeV with the excited state is at $\delta=1$~MeV above the ground state.  Again, to approximate spatial response of SPI we smoothed the line of sight integral by $3^\circ$ FWHM beam. The normalizations are chosen such that they cross at $3^\circ$. The scattering cross section of the displayed results are $\sigma_{0}=2\pi(3.11,  0.51, 0.41, 0.032,  0.0007 )\times 10^{-26}$~cm$^2$ for the ihdp, Merritt, NFW,  Cembranos, and TF distributions, respectively. The results are shown in Fig.~\ref{fig:decayprofiles} (third panel).


\subsection{Steady central source, remote production scenario}  \label{subsec:scs}


Another scenario envisions a localized source isotropically emitting particles which produce positrons in interactions that take place some distance from the source.  It is natural but not required that the source be coincident with the galactic center, defined by the visible center of the baryonic matter or by the center of the dark matter halo.  The outward streaming particles may allow both baryonic or dark matter to participate in particle creation, or may emphasize one over the other.  The baryonic or dark matter possibilities in this scenario will give the same angular profile, to the extent that the baryonic and dark matter distributions are equally cuspy and the source, baryonic matter distribution, and dark matter distribution are all concentric.

The bottom panel of Fig.~\ref{fig:decayprofiles} shows angular profiles for this central source, remote production scenario in the concentric case, using the dark matter distributions given earlier.  The angular profiles are quite steeply falling, since there is a falloff both for the $1/r^2$ flux factor and for the falloff of the target matter distribution.  In this scenario, only the ihdp dark matter profile is too flat to match the data.  Positron production off baryonic matter in this scenario should have an angular profile similar to the TF model, since this by itself gives a flat rotation curve, as is observed down to fraction of a kiloparsec.

We should expect that the central source has a finite if small extent and that the positrons will migrate at least a little before annihilating.  Accordingly, we have further smoothed the profiles by a small amount (of scale about $10^{-5}$ kpc).  This smoothing controls a mild singularity affecting some of the curves, but if the scale of the smoothing is small it does not visibly affect the curves shown in Fig.~\ref{fig:decayprofiles}.  The question of how longer positron migrations affect the angular profiles is discussed in the next major section.

A possibility within the central source, remote production scenario is that the source is separated from the baryonic matter and/or dark matter centers.  (There are examples of ``dual active galactic nuclei''~\cite{Comerford:2009ju} and of galaxies with bright off-nuclear X-ray sources~\cite{Jonker:2010ip}.)  This leads to an asymmetric or ellipsoidal distribution of positrons, as has been reported from observation in~\cite{Weidenspointner:2008zz}, as there will be two positron concentrations, one at the source and one at the matter concentration.  The hot spots may not be of equal strength, ond one or both could be muted by the $3^\circ$ FWHM detector resolution smearing.

\begin{figure}[ht]
\begin{center}

\includegraphics[width = 3.0 in]{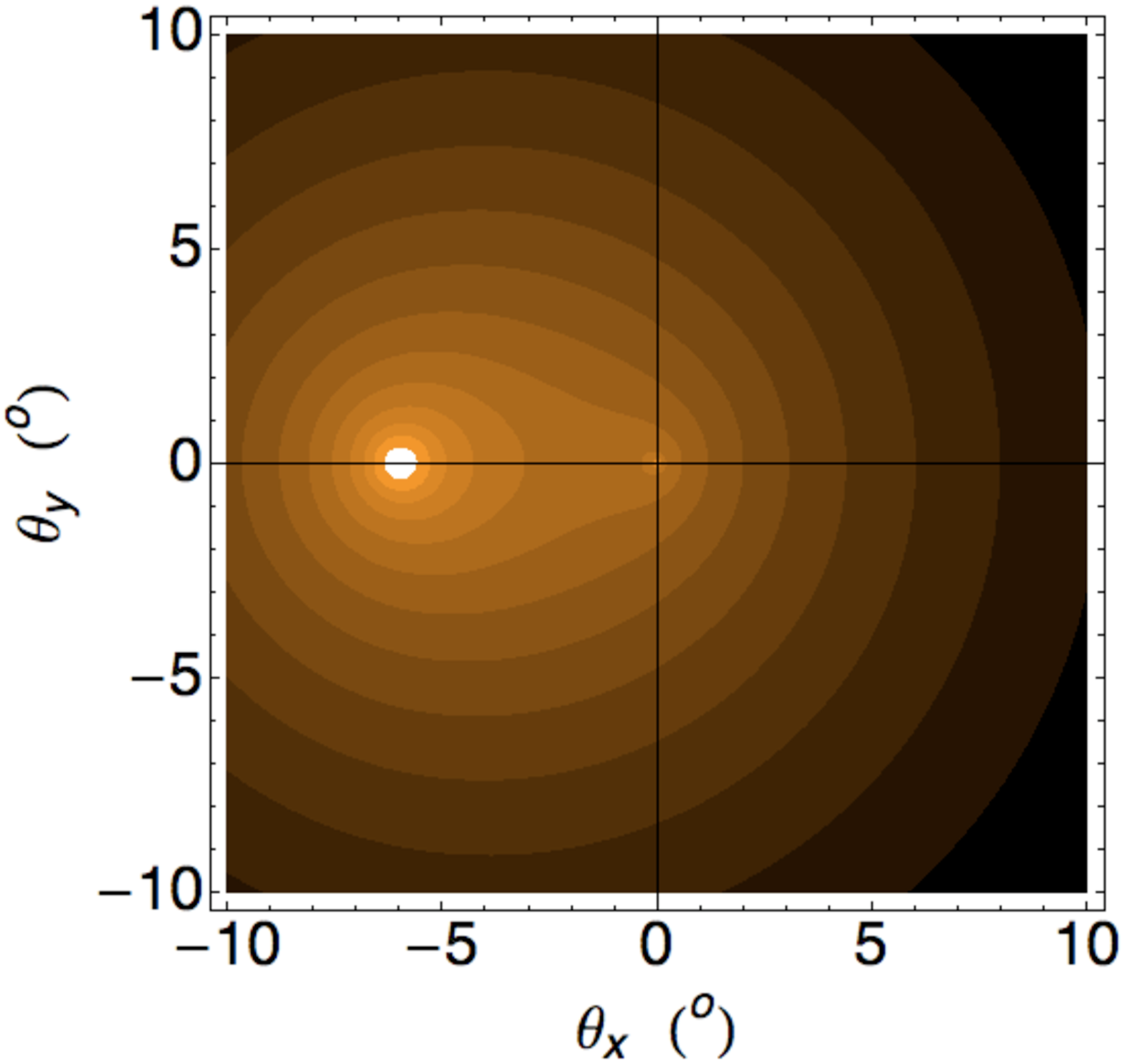}

\includegraphics[width = 3.0 in]{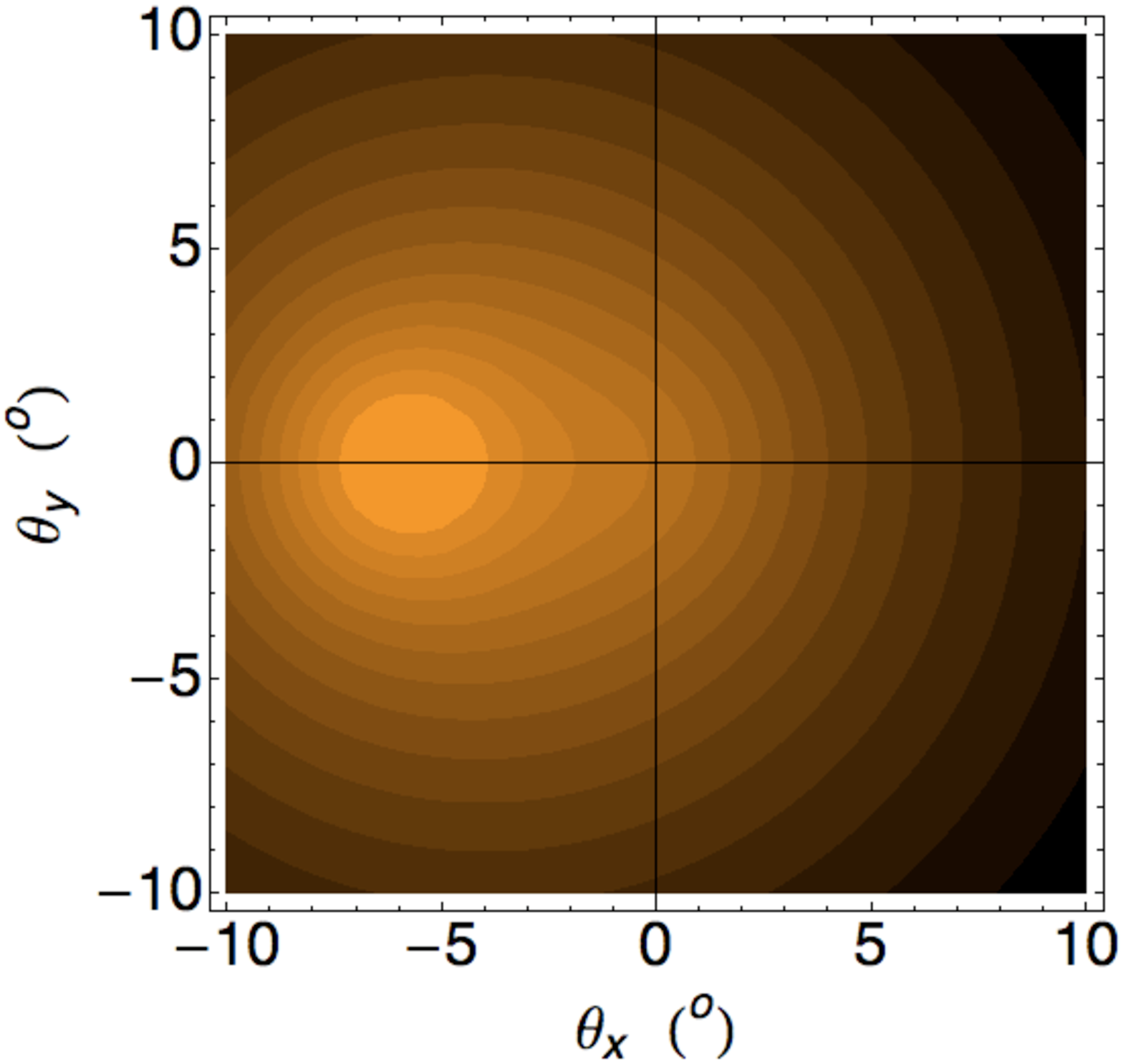}

\caption{$551$ keV gamma-ray distributions created by particles emitted from a localized source and interacting in the interstellar medium to produce positrons, in the case that the source and the mass center of the medium are separated by $6^\circ$ in the sky.  The coordinates are centered on the mass distribution, the  source is to the left.  The upper panel is unsmeared, the lower panel includes a $3^\circ$ FWHM gaussian smearing to mimic the resolution of the telescope of the INTEGRAL spectrometer.  The contours are at constant steps in the log of the intensity.}
\label{fig:point}
\end{center}
\end{figure}

We show in Fig.~\ref{fig:point} a dark matter only example of results possible in this case.  In this example, the source is taken to be the same distance from us as the relevant mass centers, but is offset $6^\circ$ in the sky from the dark matter center.  
The upper panel shows the results unsmeared.  The coordinates are centered on the mass center, and there is a bright spot there with a brighter spot at the location of the source.  The lower panel shows the results with the $3^\circ$ FWHM telescope resolution averaging.  The dark matter center hot spot is largely smeared away.   One might hypothesize within this scenario that the faint hot spot seen in the Ref.~\cite{Weidenspointner:2008zz} data could be the location of a dark matter distribution center or of a dark matter clump.  

\subsection{WIMP mass effects and extended halo effects}


The mass of the WIMP does not affect the angular profile from the dark matter relic or velocity-independent collisional scenario.  In these cases the only effect of the mass is to change the number density for a given mass density, and this is merely a constant overall factor.  

\begin{figure}[t]
\begin{center}
\includegraphics[width = 3.0 in]{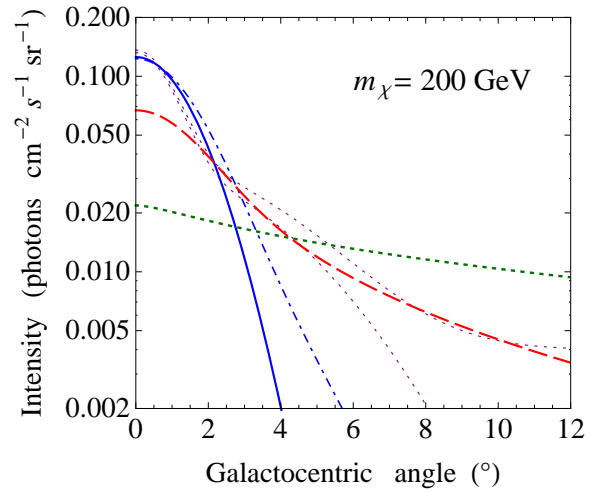}
\caption{Comparing positron production mechanisms and data, but with $m_\chi=200$ GeV. The dashed and dotted curves are the same as Fig.~\ref{fig:one} and calculated curves use the NFW dark matter distribution. The solid curve is for the excited dark matter collisional mechanism and still uses $r_{max} = 25$ kpc, while the dot-dash curve has $r_{max}=80$ kpc.}
\label{fig:rmax80}
\end{center}
\end{figure}

With a velocity dependence, mass changes can affect the shape of the angular distribution.  For the threshold model we have considered, the mass gap $\delta$ may be chosen to be just large enough to produce positrons, so that we think of $\delta$ as a constant.  Then for lighter WIMP masses, the threshold velocity increases.  Since at a given radius there is a given upper limit (the escape velocity) to the initial state velocities, there are radii where the reaction cannot proceed at all.  As the threshold velocity increases, the region where the reaction can proceed shrinks to a small ball about the galactic center.  Since we smear with a $3^\circ$ FWHM gaussian to match the telescope resolution, there is a limit to how narrow an angular profile we can obtain, and $m_\chi \approx 200$ GeV  is light enough to give a result close to this limit for the $\delta$ we have chosen.  This result is shown by the solid line in Fig.~\ref{fig:rmax80}.  Four of the other lines in this fugure are reference lines which are the same as Fig.~\ref{fig:one}.  The other new line in Fig.~\ref{fig:rmax80} is the medium weight dot-dash line, which is the $m_\chi = 200$ GeV result for $r_{max} = 80$ kpc.

\section{Positron Migration}			\label{sec:migration}


We have so far assumed that positrons annihilate near where they are made.  Positrons can migrate, for distances depending on their initial kinetic energy and the topology of the magnetic fields.  Most positrons with a few MeV initial kinetic energy traveling through matter densities typical of the galactic center will survive a few times $10^6$ years before stopping and annihilating from rest~\cite{Beacom:2005qv,Jean:2009zj}.  The energy loss, at these energies, is mainly due to ionization and collisional losses. The positrons freeze into the magnetic field lines, traveling in tight spirals around them and turning with the field lines as long as the turning radii are wider than the cyclotron radii (as is usually the case).  The actual distance the positron travels from the source hence also depends on the topology of the magnetic field.  If the field has a significant random component with a cell size of a parsec, as seems to be the case in the solar neighborhood~\cite{Sun:2007mx}, the few MeV positrons will travel about a kiloparsec before annihilating.  For the galactic center, there is progress in understanding the magnetic field~\cite{LaRosa:2004xe,LaRosa:2005ai,Morris:2007jk,Crocker:2010xc}, but it is not yet a clearly understood subject.  

The positron migration, following random trajectories from the point of their creation spiraling along random field lines before finally stopping, will be treated as a random walk and approximated by Gaussian distribution. A simple way to take this into account in the calculation of gamma ray angular profile is by performing an additional 3-dimensional Gaussian smoothing. The effect of this is a flatter gamma ray angular distribution.

\begin{figure}[ht]
\begin{center}
\includegraphics[width = 3.0 in]{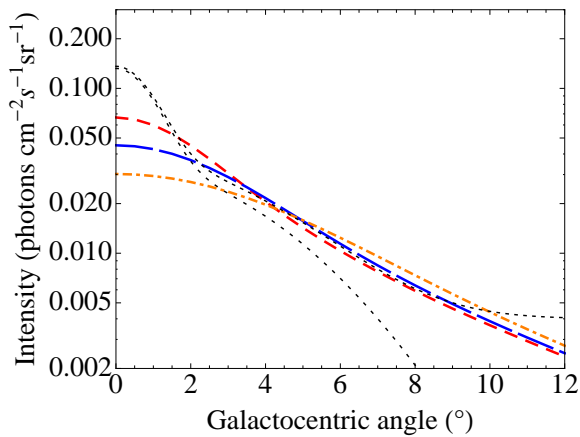}

\caption{Angular distribution for Meritt dark matter profile in the collisionally excited dark matter scenario.  Red-dashed line, blue-long-dashed line and orange-dotted-dashed line are the angular distribution corresponding to  positron diffusion length of 0.1, 0.3, 0.5 kpc FWHM, respectively.}
\label{fig:migration}
\end{center}
\end{figure}

In Fig.~\ref{fig:migration}, we show the gamma ray angular profile in the  collisionally  excited dark matter scenario with Merritt dark matter profile for three different diffusion lengths: $(0.1, 0.3, 0.5)$ kpc  FWHM.  The angular distribution in the case of $0.1$ kpc diffusion length is practically indistinguishable from the unsmoothed case in Fig.~\ref{fig:decayprofiles} and in the case of $0.3$ kpc diffusion length we still have a moderately good  fit.  (For simplicity, the width of the Gaussian smoothing used in the calculation is constant, neglecting the possibility that the diffusion length may depend on the densities of matter in the surroundings, the positron's initial kinetic energy, and the topology of the magnetic field.)

In conclusion,  positron migration may affect significantly the angular profile of the $511$ keV gamma ray from the galactic center. A  kiloparsec diffusion length appears to give a too flat profile. Indeed, with the Merritt dark matter distribution, a diffusion length of a half kiloparsec already seems to spread the angular profile unacceptably.   A large diffusion length also tends to blur the distinction among different dark matter profiles for the shape of of the gamma-ray angular distribution.


\section{Transient galactic activity scenarios}		\label{sec:sporadic}


The galactic center is currently not active and the massive black hole that lies there radiates at about 8 orders of magnitude below the Eddington luminosity~\cite{Su:2010qj}.  There is evidence that the galactic center has been more active in the past;  a few recent references include~\cite{Ponti:2010rn,Chernyshov:2009ie,Revnivtsev:2004bi}.  
Sporadic high luminosity events from the galactic center appear to have occurred in the past few hundred years~\cite{Ponti:2010rn} and there is speculation about sizable sporadic activity millions of years in the past also.  There have been considerations of how transient galactic center activity could create the positron excess~\cite{Totani:2006zx}. 

We will also consider how transient activity can connect to the galactic center positron excess.  Suppose that there have been several catastrophic events in the galactic center. The catastrophic events may be either pointlike positron injection in the center or illumination by neutral particles from a localized source, with a $1/r^2$ flux falloff, later producing positrons via interactions with the ambient matter.  

In the localized positron injection case, and we suppose positron injection occurs only at definite times, and the positrons diffuse.  Diffusion proceeds as a random walk, and the positrons form a gaussian distribution with a width that depends on the number of cells and the step size.  The width, the step size (cell size of the random magnetic field), and the number of steps (time from the positron injection) form a trio where if we know two we can work out the third.   For example, for the wider of the two gaussians in the data, the FWHM is about a kpc, and if the cell size is 1 parsec, the half-width is equivalent to 500 steps in a straight line or that number squared in the random walk.  This puts the time since the injection at a quarter million parsecs converted to time, or about 800,000 years.   The narrower gaussian is about a third as wide, so its initial injection would have occurred about $80,000$ years in the past.

The spatial profile in this scenario is described by the sum
\begin{equation}
\frac{d\Phi}{d\Omega} = \sum A_i e^{-d^2 / 2\sigma_i^2}
\end{equation}
where $d$ is the distance the positrons are from their starting point, $a$ is the step size of the random walk, $\sigma_i = \sqrt{c t_i a}$, and $t_i$ is the time elapsed since injection.  If ``$FWHM$'' is given in distance units, then the time since injection can be calculated from $t_i = FWHM^2/(8 c a \ln{2})$.  Thus a localized injection can precisely reproduce the observed angular profile of the non-disk part of the positron annihilation flux.

The coefficients $A_i$ determine the total annihilation flux.  They are determined by the number of positrons thermalized since injection.  One naturally expects a larger fraction of in-flight annihilations originating from the younger, or narrower gaussian, positron population.

In the second version, a transient localized source illuminates the
entire volume, possibly with energetic photons, that in turn produce positrons from ambient particles. The difference between this version and the ``central source, remote production'' scenario in Sec.~\ref{subsec:scs} lies in the variability of the source.  Here the galactic center region is illuminated over a time scale of 1000's of years, \textit{i.e.,} over a very short time scale compared to diffusion times.   Afterwards diffusion can occur, resulting in broadening of the positron distribution.

\begin{figure}[htbp]
\begin{center}
\includegraphics[width = 3.0 in]{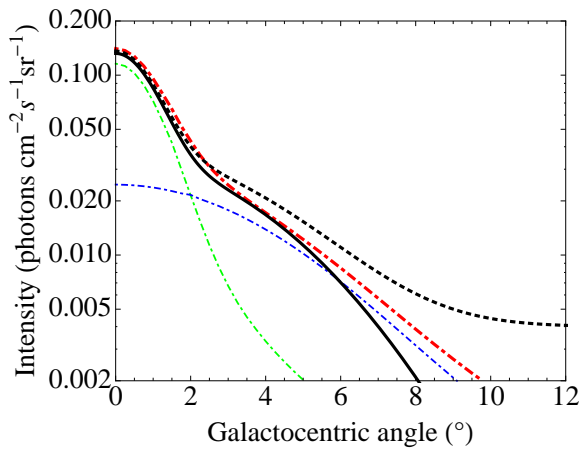}
\caption{Positron annihilation profile caused by a transient central source. The black solid curve is for centrally localized positron injection, over two different injection times.  The red heavier dot-dashed curve represents positron production from ambient Merritt \textit{ et al.} dark matter triggered by a central radiation source, also at two separate injection times indicated by the lighter dot-dashed curves.  Isotropic diffusion is applied to both cases.  The green narrower dot-dashed curve corresponds to a more recent injection event, while the broader blue dot-dashed curve curve corresponds to an older event. The dotted curve is a representation of the data with a constant piece added to indicate the level of the disc background.}
\label{fig:transient}
\end{center}
\end{figure}

Results of the two versions of this transient source scenario are shown in Fig.~\ref{fig:transient}.  Localized positrons injected at two distinct times give gaussian spatial profiles and hence, as shown by the solid curve in Fig.~\ref{fig:transient} match any gaussian representation of the data directly, with gaussian parameters determined in terms of positron diffusion characteristics.  (The dotted curve is again a representation of the data with a constant piece added to indicate the level of the disc background.) The transient localized neutral particle source with remote positron production is shown as the heavier dot-dashed curve and is made from two contributions.  One, shown as the light dot-dashed curve that is higher near the origin, has just the falloff given by the $1/r^2$ flux dependence and the density falloff of the Merritt \textit{et al.} matter distribution, and the $3^\circ$ FWHM detector resolution smearing.  The other, shown by the light dot-dashed curve that is higher at wide angles, is the same except that it has a broader spread implemented with a $8^\circ$ FWHM gaussian smearing.  These contributions are interpreted as due to one very recent outburst and one outburst farther in the past, about 400,000 to 1.6 million years ago using a cell size of $1/2$ to $2$ parsecs.


\section{Summary and Discussion}


We have examined the angular distribution of the 511-keV gamma-rays from the galactic center, mainly in the context of dark matter initiated positron  production and for a variety of models for the distribution of dark matter in the inner regions of the galaxy.  

One of the putative mechanisms, decay of dark relic particles into final states containing positrons, is not viable in light of the flatness of its angular profile compared to the data.  The degree of flatness differs for different dark matter distribution model we have examined, but in all cases the data is steeper still.

Positrons produced from collisions rather than decay show a steeper falloff with angle, mainly because of the quadratic dependence on dark matter density, and can match the observed angular falloff in the galactic bulge.  Additionally, a generic model where initiating particles steadily radiating from a central source produce positrons some distance out also gives a sufficiently steeply falling angular distribution.  Within the latter scenario, if the radiation source is displaced from the center of the mass distribution, then it leads to an azimuthally asymmetric annihilation flux.  

At large angles, the dark matter scenarios generally fall well below the data.  However, there are also definitely existing astrophysical sources of positrons in the form of, for example, low mass X-ray binaries, pulsars, and $\beta^+$ emitting nuclei from supernova explosions~\cite{Lingenfelter:2009kx}.  The numbers of supernova produced $\beta^+$ emitters, in particular, are about the right magnitude to match the large angle (and not circularly symmetric) data.

We also considered transient activity within the galactic center, characterized by several catastrophic events in the past follow by relatively quiet periods.  Using a random walk model for positron propagation we relate the spatial distribution parameters of $511$-keV emissions to characteristics of the random components of the magnetic fields in the galactic center environment.   Under this scenario, the observed gaussian spread allows us to determine products of the random field cell size and the time elapsed since the injection events.  We further considered the possibility that a transient localized source triggered remote production of positrons which then diffuse.  The resulting $511$-keV annihilation profile is broadened by an amount dependent on the time since the injection event.

Within the context of the steady state scenarios, we considered relic decay, collisional, and localized source scenarios where the positrons annihilated close to where the were formed and also considered effects of isotropic positron diffusion upon the angular profile.  (We did not consider long range or anisotropic positron transport, which others have suggested as mechanisms to alter the correlation between production and annihilation regions~\cite{Prantzos:2005pz,Higdon:2007fu,Lingenfelter:2009kx}.)  A  kiloparsec diffusion length appears to give too flat a profile.  For the Merritt \textit{et al.}~\cite{Merritt:2005xc} or the NFW~\cite{Navarro:1996gj} dark matter profiles, even a half kiloparsec diffusion length appears to lead to angular distributions that could not be reconciled with observation.  In addition, a large diffusion length tends to blur the distinction among different dark matter profiles regarding the shape of of the gamma-ray angular distribution.


\begin{acknowledgments}

ZA and CEC thank the National Science Foundation for support under Grant PHY-0855618 and AA thanks the U.S. Department of Energy for support under U.S. DOE Contract No. DE-AC05-06OR23177.

\end{acknowledgments}


\bibliography{darkmatter3}

\end{document}